# Polynomially Correlated Knapsack is NP-Complete

Chinmay Karande[*]

## Abstract

0-1 Knapsack is a fundamental NP-complete problem. In this article we prove that it remains NP-complete even when the weights of the objects in the packing constraints and their values in the objective function satisfy specific stringent conditions, *viz.* the values are integral powers of the weights of the objects.

## 1 Introduction

A general 0-1 knapsack decision problem can be formulated as the following integer program:

**Optimization Problem**

$$\max \sum_{i=1}^{n} p_i x_i$$
$$\sum_{i=1}^{n} a_i x_i \leq b$$
$$x_i \in \{0,1\}$$

**Decision Problem**

$$\sum_{i=1}^{n} p_i x_i \geq q$$
$$\sum_{i=1}^{n} a_i x_i \leq b$$
$$x_i \in \{0,1\}$$

Let us denote the decision problem above as $\mathcal{X}(A, b, P, q)$ where $A$ is the vector of weights $a_i$'s and $P$ is the vector of values $p_i$'s. In this article, we assume all input parameters to be positive integers, but our results hold true for parameters drawn from positive rationals as well. This can be proved by multiplying the equations by LCM of the denominators of the rationals involved, which yields an equivalent 0-1 knapsack problem with integer coefficients, with at most polynomially larger input.

Correlated knapsack problems are restricted cases where values are a fixed function of weights. A multitude of such problems have been studied in literature, including weakly correlated, strongly correlated and inverse strongly correlated knapsack problems [Pis98], [SYC04].

In this article, we consider another class of correlation between values and weights of knapsack. Let us define $\mathcal{X}_m(A, b, q)$ to be the following special case of the 0-1 knapsack problem for some fixed positive integer $m$:

$$\sum_{i=1}^{n} a_i^m x_i \geq q$$
$$\sum_{i=1}^{n} a_i x_i \leq b$$
$$x_i \in \{0,1\}$$

---

[*]Georgia Institute of Technology, Atlanta. `ckarande@cc.gatech.edu`

We will call this class of problems *Polynomially Correlated Knapsack problems*. Without loss of generality let us assume $a_1 \leq a_2 \leq ... \leq a_n \leq b$. Let $d_i = a_i - a_1$. Therefore, $0 = d_1 \leq d_2 \leq ... \leq d_n$.

For a solution vector $X$, we will denote by cardinality of $X$ the quantity $\sum_{i=1}^{n} x_i$, *i.e.* the number of items selected to be included in the knapsack.

Observe that $X_1(A, b, b)$ is the subset sum problem:

$$\sum_{i=1}^{n} a_i x_i \geq b$$
$$\sum_{i=1}^{n} a_i x_i \leq b$$
$$x_i \in \{0, 1\}$$

Subset sum is a well-known NP-complete problem [GJ79]. This will be the starting point of our reduction to $\mathcal{X}_m$. The rest of this article proves the following theorem:

**Theorem 1.1** *For every positive integer $m$, $\mathcal{X}_m$ is NP-complete.*

The result is interesting for the following reason: As $m \to \infty$, $\mathcal{X}_m$ becomes *easier* to solve. For very large $m$, clearly the solution is to pack the heaviest of items in the bin, as many as possible, since their values grow incredibly rapidly.

The outline of the proof is as follows: From the formulation of the 0-1 knapsack problem, one can see that it requires us to maximize along the vector $P$, in the positive orthant given a limiting constraint along vector $A$. We know that when $P = A$, the problem is nothing but subset sum, which is NP-complete. Now, when the values $a_1, ..., a_n$ are 'close' to each other, *i.e.* $d_n \ll a_1$, the vector $P' = (a_1^m, a_2^m, ..., a_n^m)$ points to almost the same direction as $A$, hence intuitively maximizing along $P'$ in this case, must be almost as hard as maximizing along $P$.

In section 2 we will prove that the special case of subset sum where $d_n \ll a_1$ is NP-complete. Then in section 3 we will reduce such instances of subset sum into $\mathcal{X}_m$ using the idea explained above. These two together will prove theorem 1.1.

## 2 A special case of subset sum

For any positive integer $m$, let $_m\mathcal{X}_1(A, b, b)$ be the restricted case of the subset sum problem where $d_n < \left(\frac{a_1}{n}\right)^{1/m}$. Intuitively, this just means that $a_i$s are very close to each other in magnitude and the vector $P$ is very close to the median ray of the positive orthant.

**Lemma 2.1** $_m\mathcal{X}_1$ *is NP-complete.*

Let $\mathcal{X}_1(A, b, b)$ be an instance of subset sum. Let $C$ be any positive integer (technically, a function of $n$, $a_n$ and $b$) such that $C > na_n^m$ and $C > b$, but such that it is polynomial in terms of $n$, $a_n$ and $b$. For example, $C$ can be equal to $na_n^m b + 1$. Let $\mathcal{Y}_r = \mathcal{X}_1(\bar{A}, \ rC + b, \ rC + b)$ be an instance of subset sum such that $\bar{a}_i = a_i + C$. Then for integral $0 \leq r \leq n$, each $\mathcal{Y}_r$ satisfies the following two properties:

**Lemma 2.2** *Every solution of $\mathcal{Y}_r$ has cardinality $r$.*

**Proof:**
For any $X$ with cardinality less than $r$,

$$
\begin{aligned}
\sum_{i=1}^{n}\bar{a}_i x_i &= \sum_{i=1}^{n}(a_i+C)x_i = \sum_{i=1}^{n}a_i x_i + C\sum_{i=1}^{n}x_i \\
&\leq na_n+(r-1)C \leq na_n^m+(r-1)C \\
&< rC \leq rC+b
\end{aligned} \quad (1)
$$

where equation (1) follows from $C > na_n^m$.

On the other hand, for any $X$ with cardinality greater than $r$, we have:

$$
\begin{aligned}
\sum_{i=1}^{n}\bar{a}_i x_i &= \sum_{i=1}^{n}(a_i+C)x_i \\
&= \sum_{i=1}^{n}a_i x_i + C\sum_{i=1}^{n}x_i \\
&\geq (r+1)C \\
&> rC+b
\end{aligned}
$$

Hence proved. □

**Lemma 2.3** $\mathcal{Y}_r$ has a solution of cardinality $r$ if and only if $\mathcal{X}_1(A,b,b)$ has a solution of cardinality $r$.

**Proof:**

- **If part**: If $\sum_{i=1}^{n}a_i x_i = b$ for some $X$ with $\sum_{i=1}^{n}x_i = r$ then we can see below that $X$ is also a solution to $\mathcal{Y}_r$:

$$
\begin{aligned}
\sum_{i=1}^{n}\bar{a}_i x_i &= \sum_{i=1}^{n}(a_i+C)x_i \\
&= \sum_{i=1}^{n}a_i x_i + C\sum_{i=1}^{n}x_i \\
&= rC+b
\end{aligned}
$$

- **Only if part**: Let $X$ be a solution of $Y_r$ of cardinality $r$, then we claim that $X$ is also a solution to $X_1(A,b,b)$.

$$
\begin{aligned}
\sum_{i=1}^{n}a_i x_i &= \sum_{i=1}^{n}(\bar{a}_i-C)x_i \\
&= \sum_{i=1}^{n}\bar{a}_i x_i - C\sum_{i=1}^{n}x_i \\
&= rC+b-rC \\
&= b
\end{aligned}
$$

□

Lemma 2.2 and 2.3 together imply that $\mathcal{X}_1(A,b,b)$ has a solution if and only if $\mathcal{Y}_r$ has a solution for some $r$.

But for every $r$, $\mathcal{Y}_r$ is an instance of ${}_m\mathcal{X}_1$. To see this note that:

$$\begin{aligned}
\bar{d}_n &= \bar{a}_n - \bar{a}_1 \\
&= C + a_n - C - a_1 \\
&= d_n \\
&\leq a_n \\
&< \left(\frac{C}{n}\right)^{1/m} \\
&\leq \left(\frac{\bar{a}_1}{n}\right)^{1/m}
\end{aligned}$$

Therefore, we have reduced $\mathcal{X}_1(A, b, b)$ to $n+1$ instances of ${}_m\mathcal{X}_1$. To see that this indeed is a polynomial reduction, observe that since $C = poly(n, a_n, b)$, we are increasing the size of input only by at most a polynomial factor going from $\mathcal{X}_1(A, b, b)$ to $\mathcal{Y}_r$ for every $r$.

Lemma 2.1 is therefore proved.

## 3 $\mathcal{X}_m$ is NP-complete

**Lemma 3.1** *The restricted subset sum problem ${}_m\mathcal{X}_1$ can be polynomially reduced to $\mathcal{X}_m$ for all $m \geq 1$.*

Note that for $m = 1$, ${}_1\mathcal{X}_1$ is a subclass of $\mathcal{X}_m$, and hence the reduction is trivial. We provide the reduction for $m \geq 2$ below.

Let ${}_m\mathcal{X}_1(A, b, b)$ be an instance of the restricted subset sum problem. Since $m$ is fixed, without loss of generality, we can assume that $a_1 > B_m n d_n^m + m$, where $B_m = {}^mC_{\lfloor \frac{m}{2} \rfloor}$ is the largest binomial coefficient of order $m$. This follows from the proof of lemma 2.1 where we can choose $C > B_m n a_n^m + m$, without affecting the proof.

Let $r = \lfloor \frac{b}{a_1} \rfloor$ and $b' = b - ra_1 < a_1$. Then the input instance can be written as ${}_m\mathcal{X}_1(A,\ ra_1 + b',\ ra_1 + b')$.

Following lemmas together prove lemma 3.1. We omit the proofs of lemma 3.2 and 3.3 as they are almost identical to proofs of lemma 2.2 and 2.3 respectively.

**Lemma 3.2** *Every solution of ${}_m\mathcal{X}_1(A,\ ra_1 + b',\ ra_1 + b')$ has cardinality $r$.*

**Lemma 3.3** *${}_m\mathcal{X}_1(A,\ ra_1 + b',\ ra_1 + b')$ has a solution of cardinality $r$ if and only if $\mathcal{X}_1(A', b', b')$ has a solution of cardinality $r$, where the vector $A'$ is defined as $a'_i = d_i$.*

**Lemma 3.4** *$\mathcal{X}_1(A', b', b')$ has a solution of cardinality $r$ if and only if $\mathcal{X}_m(A, b, G)$ has a solution of cardinality $r$, where $G = a_1^{m-1}(ra_1 + mb')$.*

**Proof:**

- **If part**: Let $X$ be a solution of $\mathcal{X}_m(A, b, G)$ of cardinality $r$. Now suppose $X$ is not a solution of $\mathcal{X}_1(A', b', b')$. Then we have two cases:

  1. $\sum_{i=1}^n d_i x_i > b'$: In this case we have,

  $$\begin{aligned}
  \sum_{i=1}^n a_i x_i &= \sum_{i=1}^n (a_1 + d_i) x_i \\
  &= a_1 \sum_{i=1}^n x_i + \sum_{i=1}^n d_i x_i \\
  &> ra_1 + b' \\
  &= b
  \end{aligned}$$

  This is a contradiction since $X$ is a solution of $\mathcal{X}_m(A, b, G)$.

2. $\sum_{i=1}^{n} d_i x_i < b'$: In this case,

$$\sum_{i=1}^{n} a_i^m x_i = \sum_{i=1}^{n} (a_1 + d_i)^m x_i$$

$$= a_1^m \sum_{i=1}^{n} x_i + m a_1^{m-1} \sum_{i=1}^{n} d_i x_i + \sum_{k=2}^{m} \left( {}^m C_k a_1^{m-k} \sum_{i=1}^{n} d_i^k x_i \right)$$

$$\leq r a_1^m + m a_1^{m-1}(b' - 1) + m B_m a_1^{m-2} \sum_{i=1}^{n} d_i^m \quad (2)$$

$$= G - m a_1^{m-2}(a_1 - B_m n d_n^m)$$

$$< G \quad (3)$$

This is a contradiction since $X$ is a solution of $\mathcal{X}_m(A, b, G)$. Here the inequality (2) and (3) follow from the previous steps by using following upper bounds on the terms:

(a) ${}^m C_k \leq {}^m C_{\lfloor \frac{m}{2} \rfloor} = B_m$.
(b) $a_1^{m-k} \leq a_1^{m-2}$ for given values of $k$.
(c) $d_i^k \leq d_n^k \leq d_n^m$ for given values of $k$.
(d) $a_1 > B_m n d_n^m$.

- **Only if part**: Let $X$ be a solution of $\mathcal{X}_1(A', b', b')$ or cardinality $r$. We have:

$$\sum_{i=1}^{n} a_i x_i = \sum_{i=1}^{n} (a_1 + d_i) x_i$$

$$= a_1 \sum_{i=1}^{n} x_i + \sum_{i=1}^{n} d_i x_i$$

$$= r a_1 + b'$$

$$= b$$

And,

$$\sum_{i=1}^{n} a_i^m x_i = \sum_{i=1}^{n} (a_1 + d_i)^m x_i$$

$$\geq a_1^m \sum_{i=1}^{n} x_i + m a_1^{m-1} \sum_{i=1}^{n} d_i x_i$$

$$= r a_1^m + m a_1^{m-1} b'$$

$$= G$$

Therefore, $X$ is also a solution of $\mathcal{X}_m(A, b, G)$.

□

**Lemma 3.5** *Every solution of $\mathcal{X}_m(A, b, G)$ has cardinality $r$.*

**Proof:**
Consider two cases

1. Let $X$ be a solution of $\mathcal{X}_m(A, b, G)$ of cardinality greater than $r$. Then,

$$
\begin{aligned}
\sum_{i=1}^{n} a_i x_i &= \sum_{i=1}^{n} (a_1 + d_i) x_i \\
&\geq a_1 \sum_{i=1}^{n} x_i \\
&\geq (r+1) a_1 \\
&> b
\end{aligned}
$$

This is a contradiction since $X$ is a solution of $\mathcal{X}_m(A, b, G)$.

2. Let $X$ be a solution of $\mathcal{X}_m(A, b, G)$ of cardinality less than $r$. Then,

$$
\begin{aligned}
\sum_{i=1}^{n} a_i^m x_i &= \sum_{i=1}^{n} (a_1 + d_i)^m x_i \\
&= a_1^m \sum_{i=1}^{n} x_i + m a_1^{m-1} \sum_{i=1}^{n} d_i x_i + \sum_{k=2}^{m} \left( {}^mC_k a_1^{m-k} \sum_{i=1}^{n} d_i^k x_i \right) \\
&\leq (r-1) a_1^m + m a_1^{m-1} n d_n + m B_m a_1^{m-2} \sum_{i=1}^{n} d_i^m \qquad (4) \\
&= (r-1) a_1^m + (m n d_n + m) a_1^{m-1} - m a^{m-1} + m B_m a_1^{m-2} \sum_{i=1}^{n} d_i^m \\
&< r a_1^m - m a_1^{m-2} (a_1 - B_m n d_n^m) \\
&\leq G - m a_1^{m-2} (a_1 - B_m n d_n^m) \\
&< G
\end{aligned}
$$

where inequality (4) follows due to the fact that $a_1 > B_m n d_n^m + m$.

This is a contradiction since $X$ is a solution of $\mathcal{X}_m(A, b, G)$.

□

Lemma 3.2 to 3.5 prove lemma 3.1. Lemma 2.1 and 3.1 together prove theorem 1.1.

## 4 Extensions and Other Remarks

We know that by variable substitution $y_i = 1 - x_i$, the knapsack problem $\mathcal{X}(A, b, P, q)$ can be reduced into $\mathcal{X}(P, q', A, b')$ and vice versa. That is, the weights and values can be interchanged. Due to this, theorem 1.1 also holds for instances when the values are $m$'th roots of the weights, i.e., $a_i = p_i^m$.

As mentioned in the introduction, although the result has been proved for integral input, it also holds for rational input.

It is a easy to see that Polynomially Correlated Knapsack problems are only weakly NP-Complete. We can devise a polytime dynamic programming algorithm to solve them if the numbers in the input are bounded by some polynomial in the input size.

**Acknowledgements:** I thank Sanjay Mehrotra and Reuben Thomas of Northwestern University and Vijay Vazirani of Georgia Tech for bringing this problem to my notice.